# Anomalous exchange coupling in transition-metal-oxide based superlattices with antiferromagnetic spacer layers.

P. Padhan[a] and W. Prellier[a],[1] and R.C. Budhani[b]

[a] Laboratoire CRISMAT, CNRS UMR 6508, ENSICAEN,

6 Bd Maréchal Juin, F-14050 Caen Cedex, France

[b] Department of Physics

Indian Institute of Technology Kanpur

Kanpur 208016, India

Abstract

A direct correlation is seen between the coercive field ($H_C$) and the magnetic-field-dependent resistivity (MR) in $SrMnO_3/SrRuO_3$ superlattices of perpendicular magnetic anisotropy. The magnetoresistance shows a sharp jump at $H_c$ for in-plane current and the out-of-plane magnetic field. Both $H_C$ and high-field MR also oscillate with the thickness of the $SrMnO_3$ spacer layers separating the metallic ruthenate. Since the spacer in these superlattices has no mobile carriers to facilitate an oscillatory coupling, we attribute the observed behavior to the spin-polarized quantum tunneling of electrons between the ferromagnetic layers and antiferromagnetically ordered $t_{2g}$ spins of $SrMnO_3$.

---

[1] prellier@ensicaen.fr

The interlayer exchange coupling (IEC) in superlattices of 3d ferromagnetic (FM) metals and non-magnetic (NM) metals oscillates between a parallel or antiparallel alignment of the magnetization vectors of the FM layers with the increasing thickness of the NM layer[1,2]. In addition, in some superlattice systems the magnetoresistance (MR) also oscillates as the thickness of the NM layer increases, and the period of oscillations in MR matches with the period of the IEC [1]. This coupling of the magnetic moments is known to be mediated by the conduction electrons of the non-magnetic layers [3,4]. Oscillations in IEC have also been observed in metallic superlattices based on the compounds of *3d*-transition metals [5,6,7]. The IEC has also been studied in the superlattice consisting of several bilayers of FM and insulator. Toscano et. al.[8] have observed non-oscillatory decay of IEC in the FM-insulator multilayer with the increasing insulating spacer layer thickness. Similar non-oscillatory decay of the FM-insulator multilayer with the insulating spacer layer thickness has also been found from in the theoretical calculation by introducing the complex Fermi surface[9,10]. Recently however, Faure-Vincent *et al*. [11] have observed the presence of antiferromagnetic (AFM) interlayer exchange coupling with insulating spacer layer. While Liu *et al* [12] have observed oscillation of IEC in a multilayer system with the increasing insulating spacer layer thickness. These observations indicate that there are still some open questions about the phenomenon of the occurrence of IEC in the ferromagnetic superlattices with the insulating spacer layer material.

In this paper we report studies of magnetoresistance and magnetization in superlattices consisting of metallic-like ferromagnetic $SrRuO_3$ (SRO) and insulator-like antiferromagnetic $SrMnO_3$ (SMO) grown epitaxially on (001) $SrTiO_3$ (STO) substrates. Our investigations reveal oscillations in the magnetoresistance (MR) and the switching field of MR and / or magnetization with the increasing SrMnO$_3$ layer thickness ($t_{SMO}$). This phenomenon is shown to be related to the interlayer exchange coupling between the SRO layers.

A multitarget pulsed laser deposition system was used to grow thin films and superlattice structures of SrRuO$_3$ and SrMnO$_3$. The details of optimized deposition conditions and structural characterization of these periodic structures are described elsewhere [13]. The superlattices were grown on (001) oriented STO substrates by repeating 15 times the bilayer consisting of 20 unit cells (u.c.) thick SRO and n unit cell thick SMO, with n taking integer values from 1 to 20. In all superlattices, SRO is the bottom layer and the multilayer is capped with a 20 u.c. SRO film to protect structural degradation of SMO. The electrical transport and magnetization measurements were performed in an external magnetic field applied along the [100], and [001] directions of the substrate. These measurements were carried out by cooling the sample to a desired temperature (T) from room temperature under zero-field conditions.

SrRuO$_3$ is a metallic ferromagnet with a Curie temperature ($T_C$) of ~ 160 K in its bulk form [14]. In contrast, SrMnO$_3$ is an antiferromagnet of Néel temperature ($T_N$) close to 260 K [15]. The $T_C$ of SRO in these superlattices is influenced by the thickness of the SMO layer.[16] The $T_C$ extracted from the field-cooled (FC) temperature-dependent magnetization of the superlattices with different $t_{SMO}$ is shown in Fig. 1. For the superlattice with 1 u.c. thick SMO layer, the $T_C$ is lower than the $T_c$ of bulk SRO[14]. As the $t_{SMO}$ increases, the $T_C$ first drops and then reaches a constant value for the higher $t_{SMO}$. The initial drop of $T_c$ can be attributed to an increasing degree of lattice strain on SRO structure due to the proximity of SrMnO$_3$. This strain eventually reaches saturation when $t_{SMO}$ exceeds a certain value. This conclusion is based on the fact that the Curie temperature of SrRuO$_3$ drops on replacing Sr$^{2+}$ with Ca$^{2+}$ ion which has a smaller radius. We also note that the magnetic easy axis of the SRO in these superlattices remains along the out-of-plane direction as in the single layer films. The superlattices show a clear saturation magnetization ($M_S$) with an enhanced coercive field ($H_C$) when measured with the external field aligned along the easy axis. For the in-plane field, the magnetization does not show any clear saturation even at 5 tesla field and the $H_C$ is much smaller in this geometry of M-H measurements. Taking into account the weak diamagnetic response of the substrate, the out-of-plane $M_S$ has been extracted by extrapolating the field linear part of the magnetic hysteresis curve (M-H) at high field, to H = 0. The resulting $M_S$ of the superlattices, recorded at 10 K, with different $t_{SMO}$ is shown in the inset of Fig. 1. The value of $M_S$ of all superlattices is lower than the $M_S$ of the bulk SRO (1.6 $\mu_B$/Ru). In this analysis we have attributed the entire ordered moment to the $Ru^{4+}$ ions and the $Mn^{4+}$ spins are assumed to be aligned antiferromagnetically. Fig. 1 shows that as the $t_{SMO}$ increases the $M_S$ of the superlattices drops rather monotonically. At the end, the $M_S$ of the superlattice with $t_{SMO}$ = 20 u.c. is reduced by a factor of ~ 3 compared to the moment of the $SrRuO_3$ film. This pronounced quenching of the ordered moments of the $Ru^{4+}$ sites in these insulating spacer based multilayers is similar to the behavior of some ferromagnetic manganite superlattices where the spacer material is also an insulator [17]. A strong correlation is seen between the (M-H) loops and the field dependence of magnetoresistance (MR) of the superlattice as seen in Fig. 2 where both MR and magnetization are shown for the $t_{SMO}$ = 1 u.c. sample. Here the magnetization shows a sudden jump to full saturation value at a critical field of ≈ ± 2 tesla. We identify this field as the switching field $H_{SW}$. The M-H loop also shows another switching with second order like transition at around H = 0 due to the presence of pin(Hard) and free(soft) $SrRuO_3$ layers[18]. A very slow initial rise of magnetization when the field is increased from zero under zero-field-condition, and also our observation of a shift of the minor loop towards positive field suggests that the magnetic coupling between the SRO layers is antiferromagnetic[18]. This inference is supported by the behavior of MR which shows a negligible variation on increasing the perpendicular field till + $H_{SW}$ is reached. At the critical field however, a sharp step-like increase in negative MR is seen followed by a field-linear MR to ≈ 7 tesla. A reversal of the magnetic field maintains a monotonic field dependence of the MR till - $H_{SW}$ is reached at which point a sharp step-like increase follows. Clearly, the positive and negative field branches of MR are mirror images of each other.

Fig. 3a shows the resistance (R) of the superlattices with n = 3 at 10 K and various values of the external magnetic field applied parallel as well as perpendicular to the film plane. The field-dependent resistance, R(H), of the superlattice with in-plane field is qualitatively similar to that of a 20 u.c. thick film of SRO. However, the R(H) curve for the perpendicular field displays a pronounced hysteretic behavior similar to the one seen in Fig. 2 for the n = 1 superlattice. In the perpendicular field direction, the field dependent resistance has both irreversible and reversible components. From a comparison with the M-H data, it is clear that the step-like drop in R at ± $H_{SW}$ is due to a switch over from AFM to FM alignment of the magnetization vectors of each $SrRuO_3$ layer. The reversible component which is monotonic in field at H > $H_{SW}$ can be identified with the gradual alignment of the pinned interfacial spins in the direction of the applied field and consequent drop in spin disorder scattering. We believe that these interfacial spins are subjected to a varying degree of pinning disorder, which makes the depinning process field-dependent.

In order to understand the effect of exchange coupling on magnetotransport, we have measured the MR at 7 tesla for the superlattices with different $t_{SMO}$. Fig. 3b and 3c show the results of these measurements recorded with field along the in-plane and out-of-plane directions of superlattices, respectively. The dependence of MR in these superlattices is strikingly different for the two orientations of the field. While the in-plane MR first increases with $t_{SMO}$ and then saturates for $t_{SMO}$ > 13 u.c., the variation of MR with $t_{SMO}$ for the out-of-plane direction of the field shows an oscillatory behavior with a peak at n ≈ 3 followed by a minimum at n ≈ 9 and a second peak with reduced MR at $t_{SMO}$ ≈ 14 unit cells. It should be pointed out here that the current in both these cases flows, on the average, along the plane of the superlattice. The MR depends on the relative orientation of the local magnetization and the spin of the mobile carriers. In the case of a superlattice, the carriers can sample, different degrees of magnetization near the SRO and SMO interface as they tunnel through the SMO layers occasionally. For the in-plane field geometry, the out-of-plane magnetization of each SRO layer tends to rotate towards the plane and a monotonic MR is expected as seen in Fig. 3(a). The smooth variation of MR with $t_{SMO}$ in this case is expected. However, for the out-of-plane field the oscillatory nature of MR with increasing layer thickness of SMO, which is an AFM insulator, is a non-trivial result. A generalization of IEC theories for insulating spacers predicts a non-oscillatory and exponentially decaying coupling as a function of the spacer layer thickness [10]. This theory however does not consider antiferromagnetism of the insulating spacer layer explicitly. In order to address this issue in some detail, we have looked at the variation of the switching field ($H_{SW}$) and coercivity ($H_c$) of the superlattices as a function of $t_{SMO}$.

Some representative data on $H_{SW}$ in the out-of-plane(H//[001]) R(H) of the superlattices for four SMO layer thicknesses are shown in the Fig. 4. The average value of $H_{SW}$ and MR are extracted from the point on the R(H) loop at which the dR/dH changes its sign from positive to negative. The resulting MR and fields ($H_C$ and $H_{SW}$) for various superlattices are plotted in Fig. 5. The variation of MR at $H_{SW}$ with $t_{SMO}$ in Fig. 5(a) shows a peak at n ≈ 3 followed by a minimum at n ≈ 9 and a second peak at n ≈ 14. The change of $H_{SW}$ and $H_C$ with $t_{SMO}$ is also qualitatively similar to that of the MR at $H_{SW}$ with $t_{SMO}$. The oscillatory dependence of MR at $H_{SW}$, $H_c$ and $H_{SW}$ on $t_{SMO}$ seen here is similar to the dependence of 7 tesla MR shown in Fig. 3c. The period of these oscillations is ≈ 11 unit cells.

In the case of metallic spacers, the period of oscillations depends on the Fermi surface parameters whereas the damping of these oscillations is proportional to the strength of the impurity scattering in the spacer. For insulating spacers, the coupling is antiferromagnetic at small thickness and decays exponentially with the thickness [10]. The coupling is also predicted to strengthen with temperature as thermally excited carriers facilitate the exchange. The insulating behavior of SMO therefore rules out any oscillatory coupling between the SRO layers. However, SMO is also an antiferromagnet with 'G' type spin ordering with alternate stacking of ferromagnetically ordered planes along the (111) direction. As in the case of CoPt-NiO-CoPt multilayers where the AFM insulator facilitates oscillatory coupling[12], it appears that the antiferromagnetic of SMO is playing a key role in magnetic exchange. The spin-polarized quantum tunneling of electrons between the $SrRuO_3$ layers[11] and the exchange interaction due to the antiferromagnetically order $t_{2g}$ spin of $SrMnO_3$ can manifest the IEC and hence the oscillation in the IEC of the superlattice system[12].

In summary, we have measured the current-in-plane magnetoresistance for both in-plane and out-of-plane magnetic fields in a large number $SrRuO_3/SrMnO_3$ superlattices. The MR for the out-of-plane field shows a sharp jump at the field corresponding to the coercivity of the superlattice. This first-time observation of a direct correlation between MR and magnetic hysteresis of an oxides-based superlattice, and the oscillatory nature of the magnetoresiatance and Hc on the thickness of the spacer is unconventional. We expect this work to stimulate further work in the theory of interlayer exchange coupling in a wider variety of superlattices where the spacer layer is an antiferromagnetic insulator.

We greatly acknowledged financial support of Centre Franco-Indien pour la Promotion de la Recherche Avancée/Indo-French Centre for the Promotion of Advance Research (CEFIPRA/IFCPAR) under Project No. 2808-1. Partial support from the European Union and the CNRS under the STREP research project CoMePhS (N°517039) is also acknowledged.

**Figures captions:**

Fig.1: Curie temperature of various superlattices extracted from the field-cooled temperature dependent magnetization with the applied field oriented along [001] direction of the substrate. Inset shows the out-of-plane saturation moment of various superlattices. The solid lines are a guide to the eye.

Fig. 2: Zero-field-cooled magnetization and MR at 10 K of the superlattice with n = 1 at different magnetic fields. The solid and dash arrows indicate the field decreasing and increasing path of the MR. The dotted lines correspond to the $H_C$ and $H_{SW}$.

Fig. 3a: ZFC resistance at 10 K of the (20 u.c.)SRO/(3 u.c.)SMO superlattices at various fields oriented along the [100] and [001] directions of STO. The arrows indicate the directions of the field sweep. The thicker arrow indicates the direction of the field at the beginning. Panel b and c show the ZFC magnetoresistance [MR=(R(H)-R(0))/R(0)] calculated from the (ρ-H) curves at 10 K of various superlattices at 7 tesla magnetic field oriented along the [100] and [001] directions of STO respectively.

Fig. 4: The ZFC resistance at 10 K at different magnetic fields oriented along the [001] direction of STO for four superlattices with n = 2, 5, 9 and 14.

Fig. 5: (a) The MR at $H_{SW}$, and (b) fields ($H_{SW}$ and $H_C$) of the superlattices with different SMO layer thicknesses at 10 K.

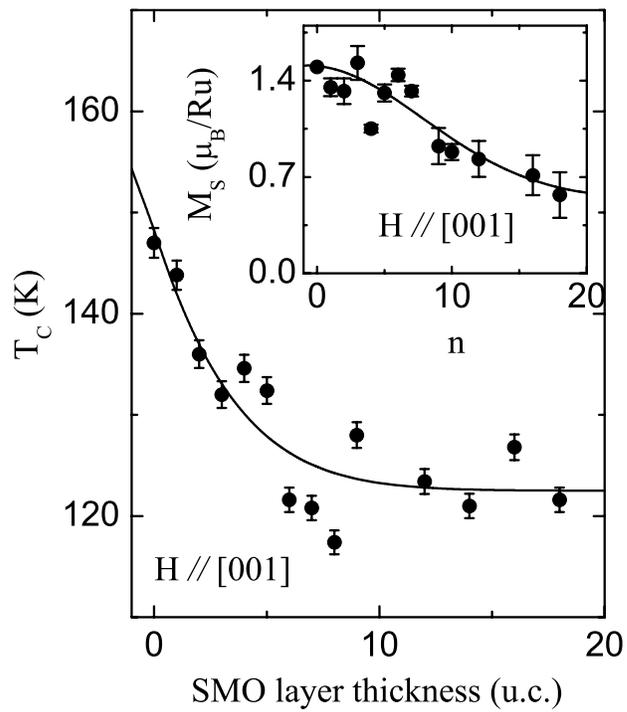

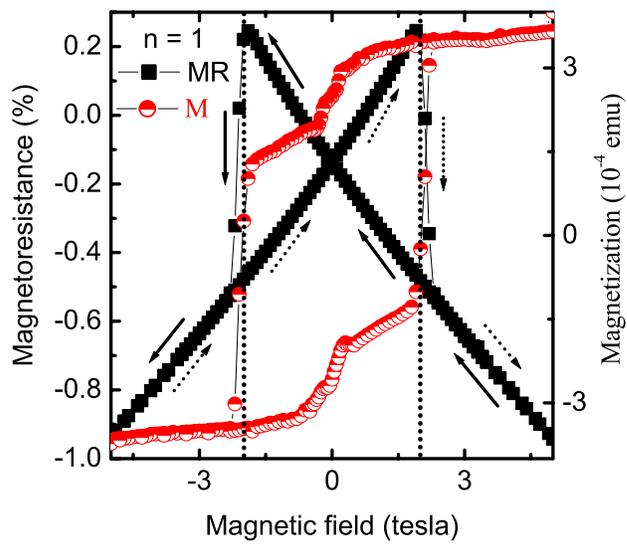

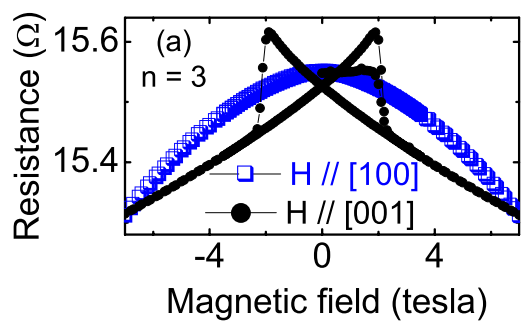
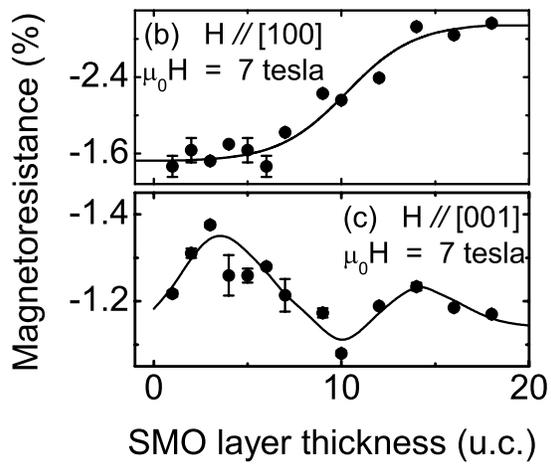

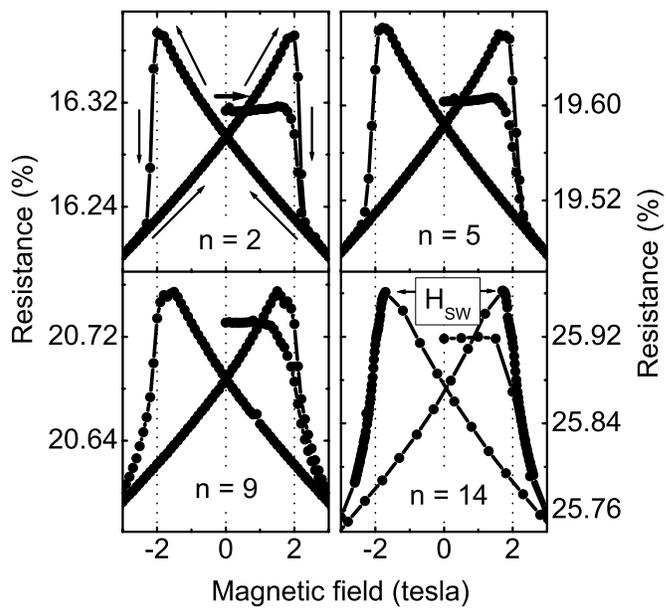

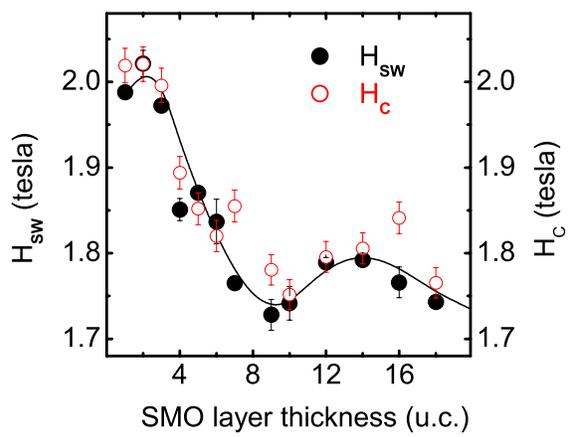